\documentclass[%
reprint,
superscriptaddress,
amsmath,
amssymb,
amsfonts,
aps,
floatfix,
]{revtex4-2}
\usepackage{graphicx}
\usepackage{dcolumn}
\usepackage{bm}
\usepackage{hyperref}
\usepackage{soul,color}
\usepackage{multirow}
\usepackage{xcolor,colortbl}
\usepackage{dcolumn}
\usepackage{mathrsfs}
\usepackage{adjustbox}
\usepackage{url}
\hypersetup{
  colorlinks=true,
  allcolors=blue,
 }
 \usepackage{array}
\begin{document}

\preprint{APS/123-QED}

\title{Performance-optimized components for quantum technologies via additive manufacturing}
 \author{S H Madkhaly}
\affiliation{
School of Physics and Astronomy, University of Nottingham, University Park, Nottingham, NG7 2RD, UK \\
}
 \affiliation{Department of Physics, Jazan University, Jazan, Kingdom of Saudi Arabia}
   \author{L A Coles}
\affiliation{
Added Scientific Ltd, Unit 4, Isaac Newton Centre, Nottingham, NG7 2RH, UK\\
}
 \author{C Morley}
\affiliation{
School of Physics and Astronomy, University of Nottingham, University Park, Nottingham, NG7 2RD, UK \\
}
 \author{C D Colquhoun}
\affiliation{
School of Physics and Astronomy, University of Nottingham, University Park, Nottingham, NG7 2RD, UK \\
}
 \author{T M Fromhold}
 \affiliation{
School of Physics and Astronomy, University of Nottingham, University Park, Nottingham, NG7 2RD, UK \\
}
 \author{N Cooper}
 \email{nathan.cooper@nottingham.ac.uk}
\affiliation{
School of Physics and Astronomy, University of Nottingham, University Park, Nottingham, NG7 2RD, UK \\
}
 \author{L Hackerm\"{u}ller}
 \email{lucia.hackermuller@nottingham.ac.uk}
 
\affiliation{
School of Physics and Astronomy, University of Nottingham, University Park, Nottingham, NG7 2RD, UK \\
}
\begin{abstract}
Novel quantum technologies and devices place unprecedented demands on the performance of experimental components, while their widespread deployment beyond the laboratory necessitates increased robustness and fast, affordable production. We show how the use of additive manufacturing, together with mathematical optimization techniques and innovative designs, allows the production of compact, lightweight components with greatly enhanced performance.
We use such components to produce a magneto-optical trap that captures $\sim 2 \times 10^8$ rubidium atoms, employing for this purpose a compact and highly stable device for spectroscopy and optical power distribution, optimized neodymium magnet arrays for magnetic field generation and a lightweight, additively manufactured ultra-high vacuum chamber. We show how the use of additive manufacturing enables substantial weight reduction and stability enhancement, while also illustrating the transferability of our approach to experiments and devices across the quantum technology sector and beyond. 
\end{abstract}

\maketitle


\section{\label{section1}Introduction}
While the growing range of quantum technologies offers great promise for both fundamental research \cite{compactinterferometer,Herrmann2012, Rosi2017} and practical applications \cite{falke2014strontium, ludlow2015optical, riehle2017optical, ExampleAtomInterferometry1, ExampleAtomInterferometry2, Blackett2, bongs2016uk}, their realization places ever greater demands on component performance. In particular, the production of portable quantum sensors \cite{knappe, salim2011compact, barrett2013mobile, schmidt2011mobile, bongs2014isense,  battelier2016development, scherer2014progress} will require compact, lightweight components capable of operating in a range of harsh environmental conditions; compactness, stability and robustness will be critical for such components and conventional, lab-based systems are not appropriate \cite{rushton2014contributed, eckel2018challenges}. 

The rapid transition of quantum technologies from research experiments to commercial devices also opens up space for innovation and the use of unconventional implementations of known techniques. 
We show how additive manufacturing (AM) can be used to create performance-optimized components, unimpeded by the constraints of conventional manufacturing methods, while at the same time allowing quick and easy production of customized components and thus greatly accelerating the prototyping and testing of novel component designs. The approach is generalizable to a wide range of experimental components and will transform applications as diverse as miniaturized optical devices, vacuum systems and magnetic field generation. Our work complements previous studies of integrated laser sources \cite{rocket2017jokarus, rocket2017autonomous, zhang2018compact} and miniaturized vacuum chambers \cite{mcgilligan2020laser} and expands preliminary studies of the utility of additive manufacturing in the setting of quantum technologies \cite{vovrosh2018additive, Reece}.
\begin{figure*}
    \centering
    \includegraphics[width=0.7\textwidth]{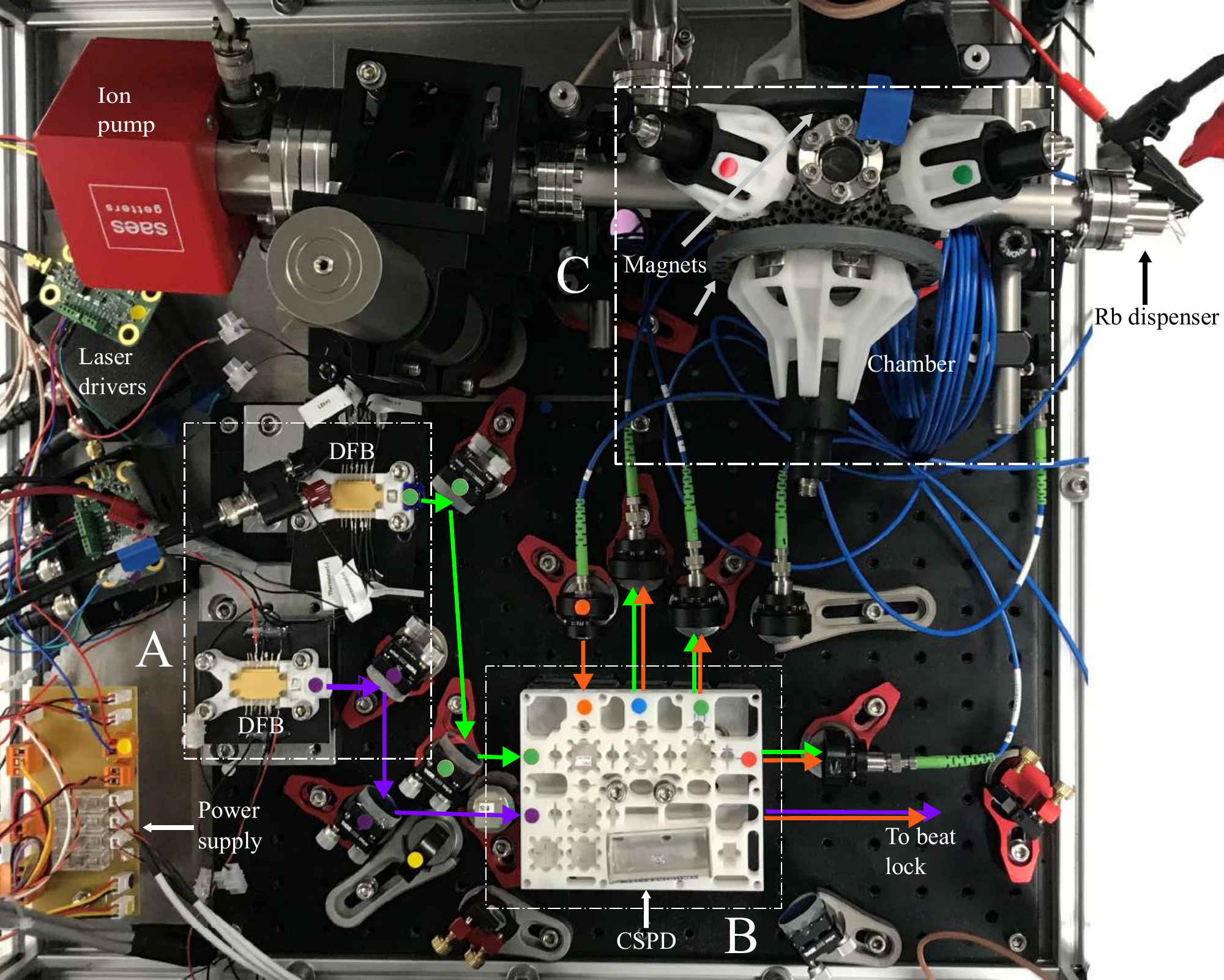}
    \caption{Overview of the complete setup, showing 3D printed and optimized components in the areas marked with dashed boxes A, B, and C. A indicates the distributed feedback lasers (DFBs) used as master light sources, B indicates the compact spectroscopy and power distribution apparatus (CSPD), and C indicates the trapping apparatus including the AM UHV chamber, optimized permanent magnet arrays and a set of self-aligning AM fiber outcoupler mounts. The setup takes up a volume of 0.15\,m$^{3}$ and the custom parts indicated have a cumulative mass of 3.2 kg.} 
    \label{optabox}
\end{figure*}
Specifically, we demonstrate a new approach to experimental design in free-space optics, where the overwhelming majority of the adjustable components are eliminated and most of the optical elements are mounted in a monolithic, additively manufactured mount within pre-aligned push-fit slots. This new approach offers improvements in stability as well as significant reductions in cost and in size, weight and power consumption (SWAP). We apply this technique to create a stable mount for an optically isolated laser source and a compact and highly stable apparatus for optical power distribution and laser frequency stabilization. 
\\
\indent The described components are combined with an AM ultra-high vacuum chamber \cite{chamberpaper} to form a magneto-optical trap (MOT), which captures 2\,$\times$\,10$^{8}$ cold $^{85}$Rb atoms. The MOT is the starting point for nearly all cold-atom based experiments and quantum technologies \cite{raabmot}. The magnetic fields required for our MOT are produced using an array of neodymium magnets in a custom-built AM mount, offering significant SWAP reductions over conventional MOT coils. An optimization algorithm was developed to determine the placement of the permanent magnets in order to accurately replicate the conventional anti-Helmholtz field used in a MOT; the algorithm is transferable to the recreation of other field structures. Our results demonstrate the power of AM to directly implement the outcome of an optimization process, without reference to traditional manufacturing constraints. Measurements of the atomic lifetime within the MOT are used to place an upper limit on the background pressure in the AM vacuum chamber of $\sim$\,3\,$\times$\,10$^{-9}$\,mbar. An overview of the MOT system is given in figure \ref{optabox}.
\\
\indent The remainder of this paper is organized as follows: the overall setup is explained in Section~\ref{section2}, including a description of the laser sources used (Section~\ref{section2.1}), followed by a discussion of the compact AM spectroscopy and power distribution system (Section~\ref{section2.2}). The placement of the ferromagnets used for MOT field generation and corresponding optimization algorithm are described (Section~\ref{section2.4}) and a brief overview of the AM vacuum chamber is given (Section~\ref{section2.3}). The performance of the MOT is characterized in Section~\ref{section3}.

\section{\label{section2}System architecture and components}
\subsection{Laser sources}
\label{section2.1}
\vspace{-0.5cm}
With the atomic structure of alkali metal atoms in mind, in particular $^{85}$Rb - see Fig.~\ref{err}(a-c), the typical roles of lasers employed for magneto-optical trapping are used here \cite{raabmot}: `reference' and `repumper' lasers, each frequency-stabilized directly to an atomic transition via saturated absorption spectroscopy \cite{SAS, SAS2}, and a `cooler' laser that is stabilized at a fixed frequency offset from the reference laser via an optical beat signal \cite{MITbeatlock}. 

\begin{figure}
     \includegraphics[width=0.45\textwidth]{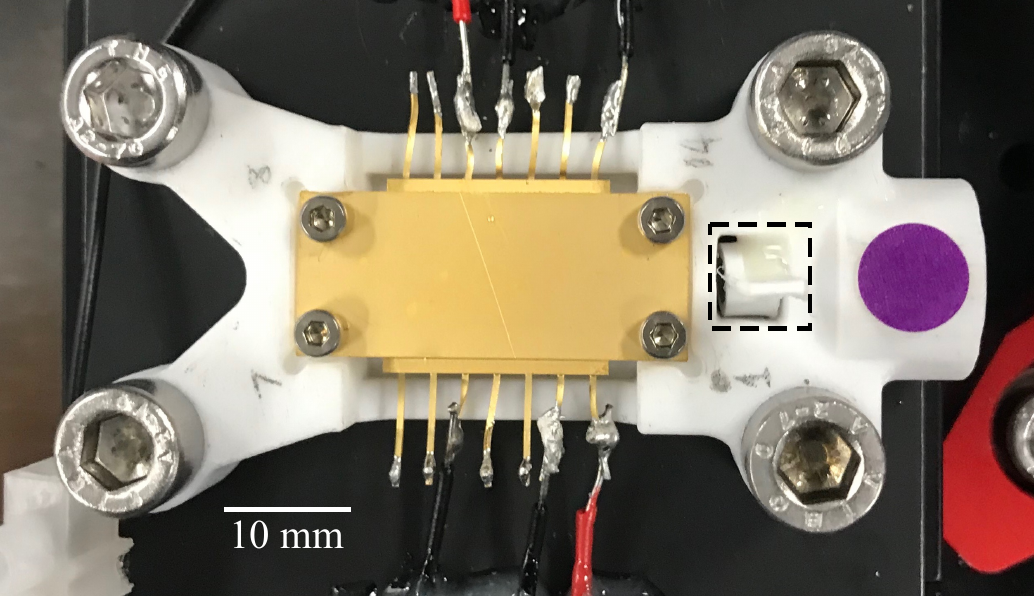} 
    \caption{Photograph of a butterfly-packaged DFB laser and optical isolator (indicated with the dashed rectangle) in an AM mount.}
    \label{dfb}
\end{figure}

Fig.~\ref{optabox}(A) shows the distributed feedback lasers (DFBs) used as our reference and repumper lasers; DFBs were chosen for their stability, large mode-hop free tuning range and compact size. Specifically an Eagleyard laser diode is used with an output power of 80\,mW at 780\,nm. Although this is sufficient to produce a MOT, a tapered amplifier (Toptica TA-100, cooler laser), which provides up to 1\,W of output power, is also used to facilitate further experiments and provides the `cooler' light for our MOT. The DFB packages are encased in an AM mount with an optical isolator (Isowave I-780-LM), as seen in Fig.~\ref{dfb}. The optical isolator has an external diameter of 4\,mm and a depth of $\sim2.9$\,mm. It is mounted inside an AM ring with a small grip to facilitate adjusting the isolation angle manually. The total weight of the laser ensemble is only 51\,g. 
After optimizing the isolation efficiency, the isolator mount was secured in the DFB housing using epoxy adhesive.  \\
\indent Figure \ref{dfb} shows the physical implementation of the corresponding systems. The AM mount leaves the back plane of the laser package exposed to permit heat sinking, so that the case can act as a thermal reservoir for the internal thermoelectric cooler used to stabilize the diode temperature.\\

The polymer casing surrounding these systems and holding the various components in place was produced from photopolymer resin (Formlabs `Rigid Resin') via stereolithographic additive manufacturing (SLA) \cite{bagheri2019photopolymerization}. SLA allows a customized mount to be produced to meet the component and alignment requirements of the user. Once the casing is printed the components simply slot into position, reducing the need for user-alignment. The photopolymer resin offers a good balance between weight and thermal/mechanical stability, --- see supplementary material for more details.
The completed reference and repumper source assemblies both provide output powers of 42\,mW, consistent with  80\,mW output power from the DFBs and 2.8\,dB losses in the isolators when used optimally.  
The DFB lasers are powered using the Koheron DRV200-A-200 compact driver board, which provides diode currents of up to 200\,mA and allows current modulation at up to 6\,MHz.
\subsection{\label{section2.2}Laser spectroscopy and optical power distribution}
\begin{figure*}
    \centering
    \includegraphics[width = 0.80\textwidth]{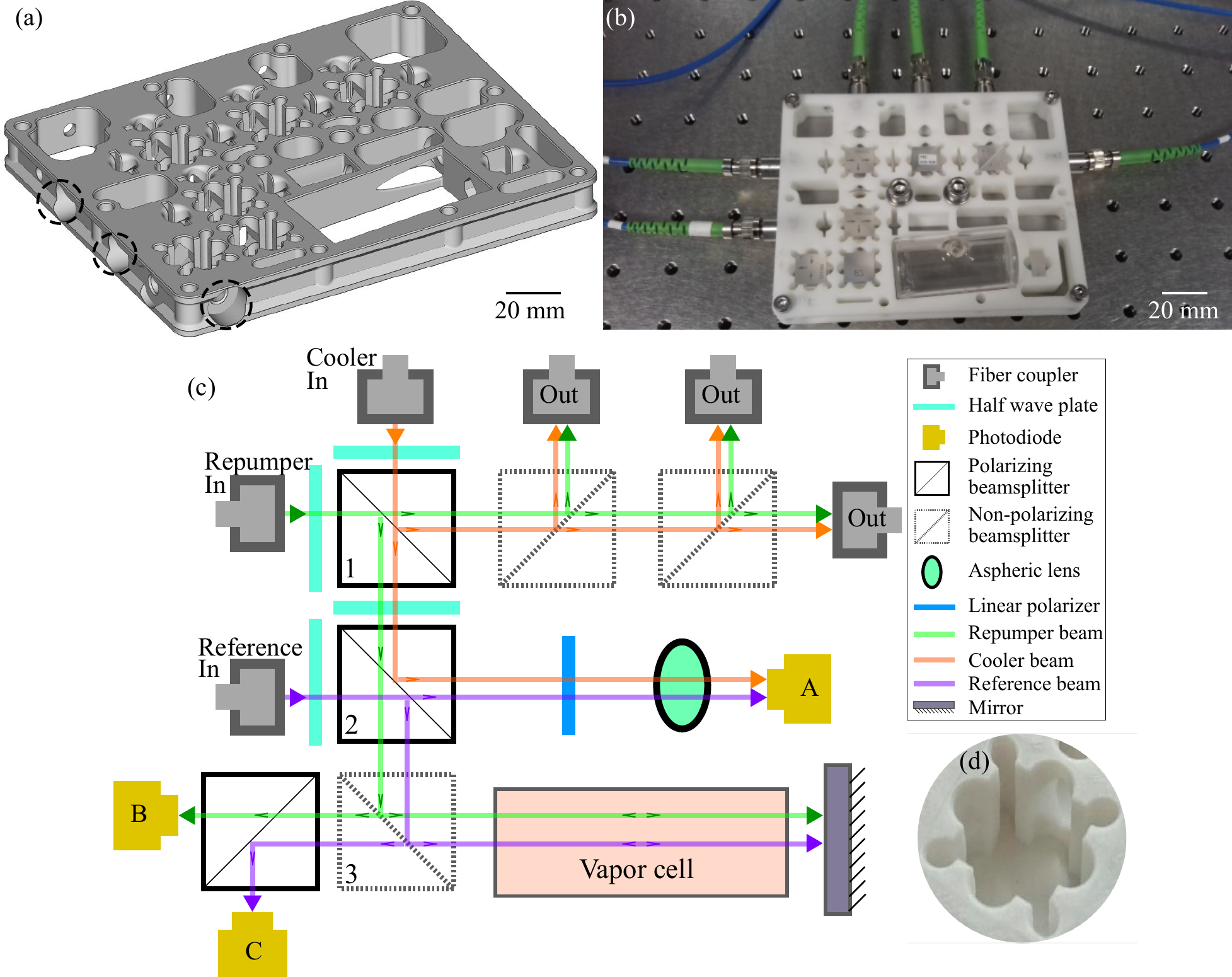} 
    \caption{The compact spectroscopy and power distribution apparatus. (a) A 3D render of the mount. The holes that can be used as fiber inputs or outputs are indicated. (b) A photograph of the CSPD with optics, a reference cell and fibers adhered to the appropriate positions. (c) A schematic of the optics layout in the CSPD, and how each input laser beam is directed through it. The purple beams represent reference light; the orange beams cooler light and the green beams repumper light. (d) Close-up of a beamsplitter slot. The rounded recesses on the edges/corners are there to prevent scuffing of the optically active surfaces and to improve push-fit alignment accuracy, respectively.}
    \label{CSA} 
\end{figure*}

\vspace{-0.3cm}
To enhance laser stability, all of the optics required for laser stabilization via vapor cell spectroscopy \cite{preston1996doppler} and optical beat locking \cite{schunemann1999simple, MITbeatlock}, as well as those needed to distribute the optical power of the cooler and repumper lasers appropriately between the MOT beams, were secured in a custom-designed optics mount with few adjustable elements. This mount, designated the `compact spectroscopy and power distribution' apparatus (CSPD), uses fixed beam paths and pre-aligned components (via push-fitting into specially designed slots in the AM mount) to eliminate the need for adjustment screws or tunable mirror-mounts. The result is a robust and stable setup --- see Fig.~\ref{optabox}(B). 
Like the laser mounts, the CSPD was manufactured from Formlabs `Rigid Resin' via an SLA process. The CSPD is shown in Fig.~\ref{CSA} and has dimensions of 128$\times$103$\times$12\,mm, and a mass of 84\,g. Rigid Resin was selected as its build material based on a thorough assessment of the physical properties of the available build materials --- see supplementary material for more details. The CSPD design uses the fewest optical elements possible, in order to enhance stability and reduce cost and SWAP.

To improve stability, the optical beam paths in the CSPD are kept as short as possible and the number of reflections undergone by each beam is also minimized. Each beam that is fiber-coupled for transmission to the MOT undergoes a maximum of 2 reflections in this arrangement, as does each of the two beams combined to produce the optical beat signal. The maximum path length for any of these beams is 120 mm. The (less alignment-sensitive) saturated absorption spectroscopy beams each undergo four reflections, with a maximum optical path length of 290 mm.  

To illustrate the importance of this, consider the following simple model of a beam path subject to experimental imperfections. The expected positional deviation of a beam from its target at the end of an optical beam path is given by
\begin{equation}
    \Delta r = \left[ \sum_{\mathrm{i}} (2 \Delta \phi L_{\mathrm{i}})^2 \right]^{1/2}, 
\end{equation}
where the sum is taken over all reflective components in the beam path, which are all assumed to have independent alignment inaccuracies of magnitude $\Delta \phi$. The values of $L_{\mathrm{i}}$ are the remaining path lengths between each component and the end of the optical beam path, and we have applied the small angle approximation $\tan{\theta} \approx \theta$. We can now compare the CSPD with a more conventional setup. In the CSPD, prior to fiber coupling the cooler beam undergoes 2 reflections with $L_1 = 30$\,mm and $L_2 = 90$\,mm, yielding $\Delta r = 190 \Delta \phi$ mm. By contrast, a more conventional system might involve say ten reflective components, roughly equally spaced along a 1\,m path length. In this case we find that $\Delta r = 3920 \Delta \phi$ mm, more than 20 times the equivalent value for the CSPD. 
Each of the optics slots within the CSPD was designed to leave clearance in the center of the optical component; this prevents direct contact between the mount and the central part of the optical element on which the beam impinges, thus ensuring that device performance is not degraded by scuffing of the optical surfaces when components are inserted. Slots for cube beamsplitters have rounded recesses in each corner (similar to an undercut) --- see Fig.~\ref{CSA}(d). This improves the accuracy of the push-fit alignment by ensuring that cube position/orientation is controlled via extended contact with defined flat surfaces that can be built accurately using the SLA process. This is important because AM methods are not well-suited to producing sharp, internal features, such as the corners of the beamsplitter slots.

The layout of the optics and optical paths within the CSPD were designed to minimize the number of optical components required. A schematic of the beam paths used in the CSPD is shown in Fig.~\ref{CSA}(c). The cooler beam first passes through a $\lambda$/2 wave plate that controls the distribution of the power between the MOT beams and the spectroscopy setup, which happens at the polarizing beamsplitter marked `1'. The reflected component is divided among the three output fiber couplers which provide the MOT beams. This happens at the two non-polarizing beamsplitters immediately to the right of polarizing beamsplitter `1'. The first of these has a splitting ratio of 67/33, while the second splits the power 50/50. The repumper beam also reaches polarizing beamsplitter `1' via a wave plate to control power distribution. In this case, the transmitted polarization component is distributed among the MOT beams.

The reflected component of the repumper beam and the transmitted component of the cooler beam then pass through a half wave plate fixed at 45$^\circ$ relative to the polarizing beamsplitters. Thus, when they are combined with the reference beam at polarizing beamsplitter `2', the cooler light is reflected and the repumper light transmitted. The component of the reference light transmitted at beamsplitter `2' is mixed with the cooler light on the same pathway by the polarizing filter, again fixed at 45$^\circ$ relative to the polarizing beamsplitters, such that an optical beat signal can be produced on photodiode `A'. This is used to stabilize the frequency difference between the cooler and reference lasers via feedback to the diode current of the cooler laser. 

\begin{figure*}
\centering
   \includegraphics[width=\textwidth]{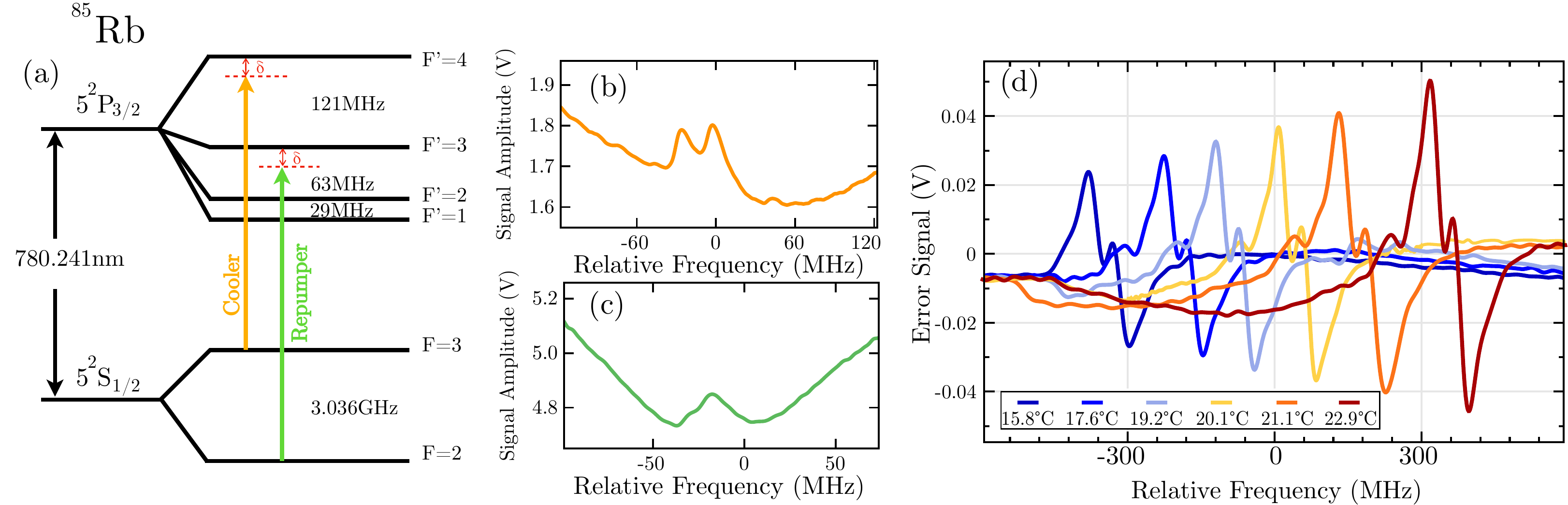} 
    \caption{ \label{err}Laser locking frequencies relative to $^{85}$Rb D$_{2}$ transitions (a), and the corresponding saturated absorption spectroscopy features of cooler (b) and repumper (c) beams. While (d) shows the corresponding error signal generated from the saturated absorption spectroscopy optics in the CSPD for the reference beam, as a function of relative laser frequency. The form of the signal can be seen to remain consistent and appropriate for feedback stabilization of the laser frequency despite substantial variations in environmental temperature. A horizontal offset has been added to improve visibility.}
\end{figure*} 
This leaves the repumper light transmitted at beamsplitter `2' and the component of the reference light reflected at beamsplitter `2'. These enter what is a conventional saturated absorption spectroscopy setup for one laser --- the difference here is that two beams overlap on the same spatial path in orthogonal polarizations. After re-emerging from the spectroscopy setup, they are ultimately separated onto their respective photodiodes by the polarizing beamsplitter `3'. To generate an error signal suitable for feedback stabilization of the laser frequencies, the laser currents are sinusoidally modulated and the modulation signals are combined with the photodiode outputs using analog multipliers --- a standard practice in laser stabilization \cite{citation-0}.

By sharing the same spatial path in the vapor cell the reference and repumper beams influence each other's spectroscopy signals via optical pumping effects \cite{smith2004role}. Provided that both lasers are to be stabilized simultaneously this does not prove detrimental to the operation of the device, but rather increases the strength of the locking signals (see supplementary material for more details). 


The design of the CSPD allows fixed-focus fiber collimators to be inserted directly into the fiber access ports (see Fig. \ref{CSA}) and fixed in place using epoxy once aligned. However, the stability of standard fiber-optic connectors was found to be insufficient to allow long term operation without adjustable components. This could be fixed by a custom fiber mount. In principle, this device can be extended to unite the laser source housing with the power distribution and spectroscopy optics, thus eliminating the need for any external optics (or an associated baseplate/breadboard) and providing all of the light generation requirements for a MOT in a single, stand-alone, fiber-coupled device.

\textit{Thermal stability and resistance tests.} - Temperature fluctuations are a major source of  drifts in optical alignment, even in a temperature-stabilized lab environment; outside the lab these problems become much more significant.
To test the thermal stability of the CSPD, the environmental temperature was adjusted between 288\,K and 298\,K, while monitoring key parameters of the system.

Figure \ref{err} shows the error signal generated for feedback stabilization of the reference laser frequency at a range of environmental temperatures within this window. Note that the results for different temperatures are intentionally offset relative to one another to improve visibility. It can be seen from the figure that the overall form of the signal is unaffected by beam misalignment due to temperature variations; it remains appropriate for laser stabilization over the entire temperature window. The change in signal amplitude occurs due to an increased vapor pressure in the reference cell at higher temperatures. 

One experimental parameter that is extremely sensitive to beam misalignment is the coupling efficiency of light into optical fibers. The optical power coupled into the fibers at the outputs of the CSPD was monitored and only 10\,\% power variation was observed over the entire 10\,K temperature window, with a maximum relative power variation coefficient of 0.02\,K$^{-1}$. This result represents a significant advance over standard lab optics and optomechanics, where typically a change of $\sim 1\,$K can lead to a complete coupling loss, and is comparable to results obtained with much heavier and more costly systems built out of materials specifically selected for thermal stability, such as Invar and Zerodur
\cite{duncker2014ultrastable}. 

The design concept behind the CSPD, i.e. the use of a monolithic, additively manufactured optics mount as a replacement for conventional optomechanics, is generalizable to almost any desired arrangement of free-space optics. Our results show that this approach offers major advantages in terms of compactness, stability, cost and assembly time. Further work in this area may lead to the establishment of a new paradigm for experimental design with free-space optics.

\subsection{\label{section2.4}Magnetic field generation}
\vspace{-0.3cm}
The efficient creation of a linear MOT field, prioritizing both field fidelity and power consumption, is an important consideration for a portable apparatus. In conventional systems, the required fields are generated by coils drawing many watts of power. The apparatus developed here instead utilizes an array of ferromagnets to generate MOT-suitable magnetic fields, thus eliminating the power consumption of the coils entirely. This technique has not traditionally been employed for SWAP reduction because many experiments require the magnetic fields to be briefly extinguished following the collection of an atomic cloud. However, we show that even in such cases it is still possible to augment MOT coils with permanent magnet arrays, and that doing so reduces time-averaged power consumption by a factor of $1/(1-T_M)$, where $T_M$ is the fraction of the experimental cycle time for which the magnetic field should be present --- see Supplementary Material for full details.  

\begin{figure}[htbp]
    \centering
    \includegraphics[width = 0.35\textwidth]{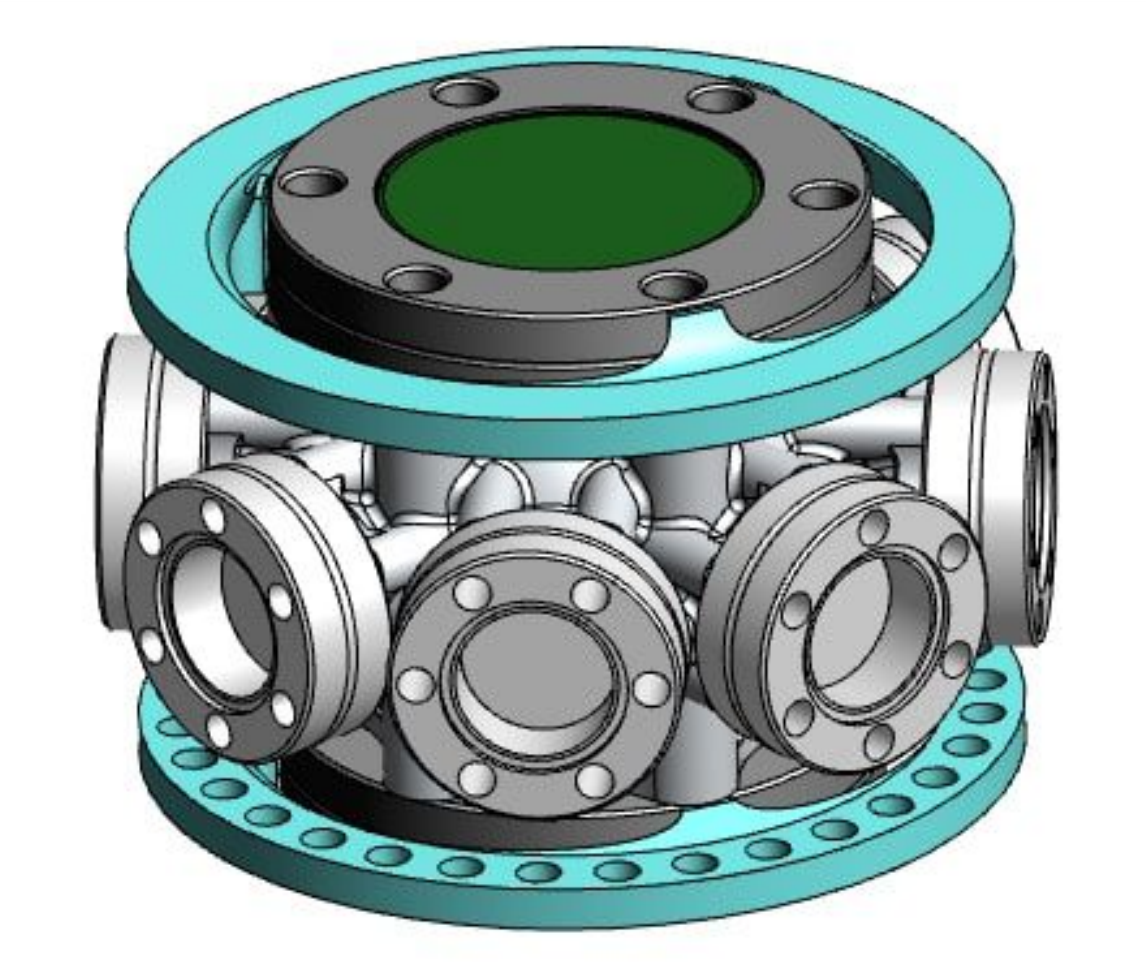} 
    \caption{A 3D model of the vacuum chamber seen in FIG.~\ref{optabox}, without the lattice structure. The blue rings attached to the top and bottom of the chamber represent the permanent magnet arrays.}
    \label{magnets} 
\end{figure}


In order to open up these experimental possibilities it is necessary to design a ferromagnetic array that produces the same field distribution as conventional MOT coils. This can be done via well established numerical optimization methods \cite{adby2013optimisation} and computer science algorithms \cite{matousek2007understanding} allowing the optimal placement of ferromagnets to be determined \emph{a priori} for any given apparatus, thereby greatly reducing testing and manufacturing times. Neodymium magnets are manufactured in a variety of standardized shapes and strengths. This makes them an ideal choice for an optimization algorithm designed to determine the optimal placement of a set of magnetized voxels on a predetermined initial grid to create a required field profile, allowing multiple magnet configurations to be designed and tested to establish their suitability before manufacture. 
\begin{figure*}
    \centering
   \includegraphics[width=0.70\textwidth]{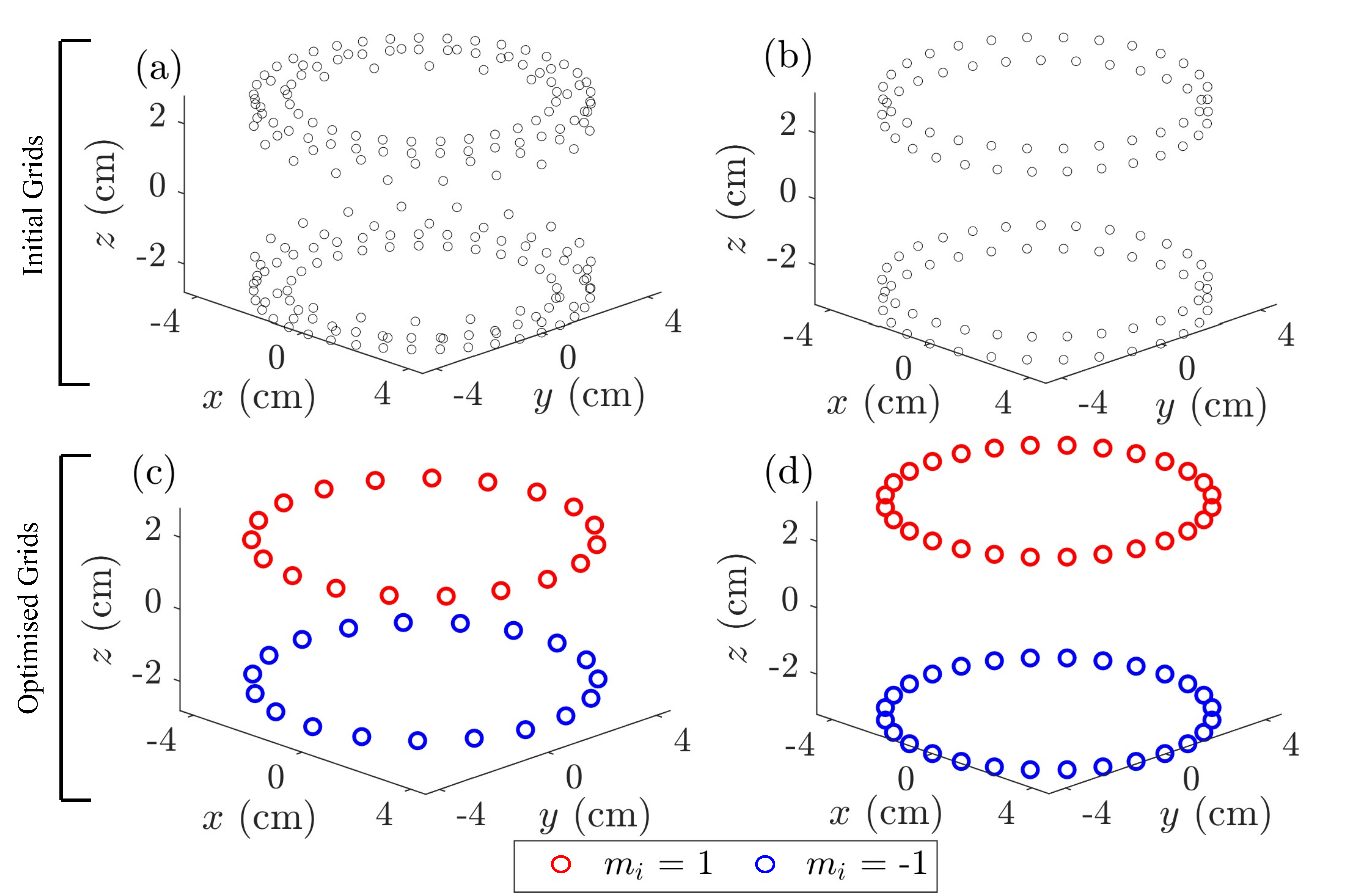} 
    \caption{Grids of the possible positions for two different types of magnetized voxels: (a) diameter of 6\,mm, depth of 3\,mm and (b) diameter of 6\,mm, depth of 6\,mm. The optimized arrangements to produce a MOT field are shown in (c) and (d) for the initial grids of (a) and (b), respectively. The geometry in (d) was used as a basis for the magnet rings shown in Fig.~\ref{magnets}, as the increased distance between the rings and the trapping region was necessary to accommodate the vacuum chamber.}
    \label{fig: Grid Optimisations}
\end{figure*}
\begin{figure*}
    \centering
    \includegraphics[width=0.75\textwidth]{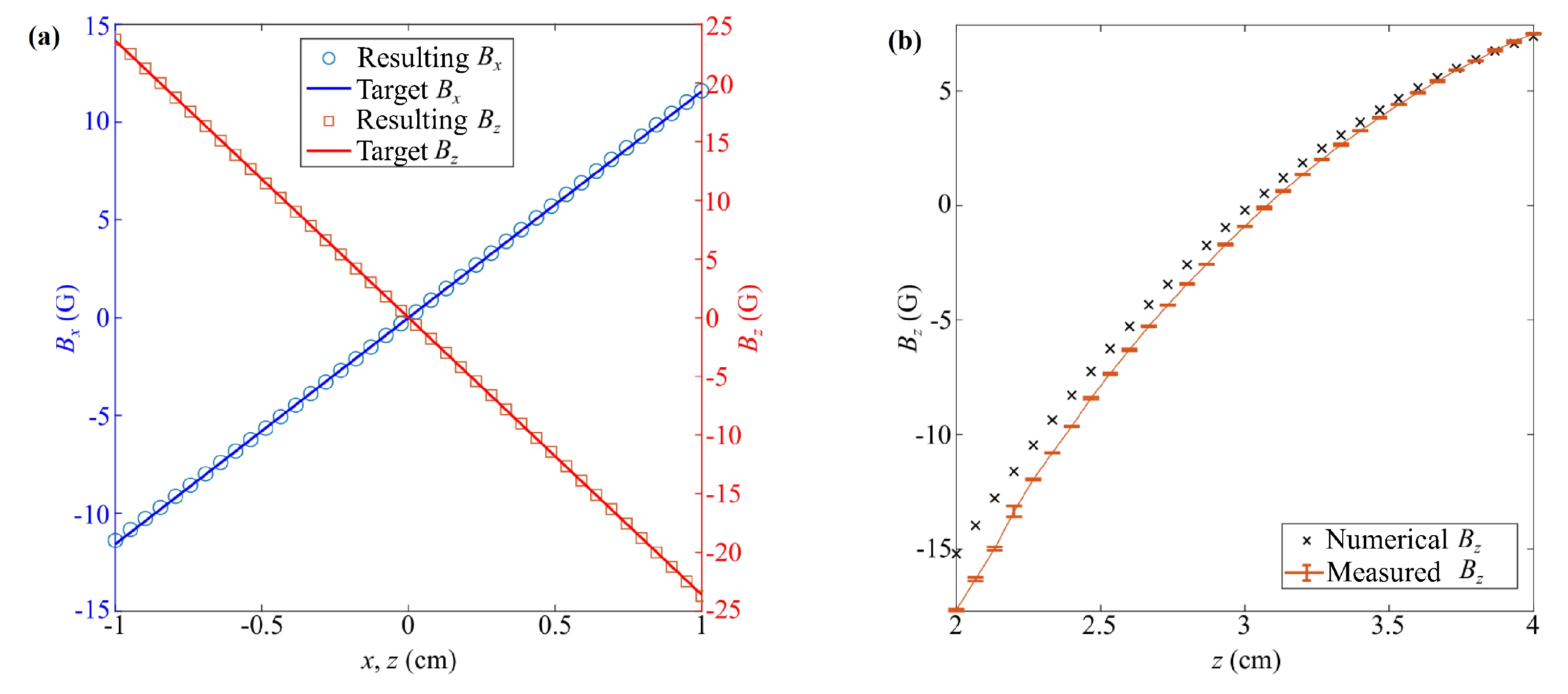} 
    \caption{(a) Graph showing the similarities between the numerically calculated fields produced by the optimized magnet structure (circles and squares) of Fig.~\ref{fig: Grid Optimisations}(d) and target magnetic fields (lines) for $B_x$ along the $x$-direction (blue) and $B_z$ along the $z$-direction (red). (b) Comparison between the numerically calculated and the experimentally measured magnetic field at the center produced by a single ring of magnetic voxels vs. distance from the ring.}
    \label{fig: Field Comparisons}
\end{figure*}
Such algorithms are frequently used in the context of magnetic resonance imaging (MRI) \cite{while20103dcoil} in the process of shimming, which employs optimization algorithms to inform the placement of permanent magnets or current carrying wires to remove unwanted spherical harmonic contributions in a given region. These techniques for passive shimming \cite{li2015mainpassive} were exploited in the design of the ferromagnet array. A voxel of volume $dV$, magnetized entirely along $z$ with strength $M_z$, at a position $\underline{Q} = (\rho,\alpha,\psi)$ produces a scalar potential at a position $\underline{P} = (r,\theta,\phi)$ given by,
\begin{eqnarray}
\Psi = -\frac{\mathrm{d}V M_z}{4 \pi \rho^2} \sum_{n = 0}^{\infty} \, \sum_{m= 0}^{n} \epsilon_m \frac{(n - m + 1)!}{(n + m)!} P_{n+1}^m \nonumber\\ (\mathrm{cos} \, \alpha) \left( \frac{r}{\rho} \right)^n P_n^m (\mathrm{cos} \, \theta ) \mathrm{cos}(m (\phi - \psi))
\label{eq:Greens Harmonics Full}
\end{eqnarray}
\noindent where $P_n^m$ are the associated Legendre polynomials and $\epsilon_m$ is the Neumann factor defined as $\epsilon_{m = 0} = 1$ and $\epsilon_{m > 0} = 2$. This can be related to the magnetic fields by $B = -\nabla \Psi$. From this, a matrix equation can be formed relating the spherical harmonic contributions from each voxel and its magnetization to the required overall spherical harmonics, 
\begin{eqnarray}
\label{eq:Spherical Harmonic Matrix} 
\begin{pmatrix}
A_{(z,1)} \, (1,0) &  \dots & A_{(z,N)} \, (1,0)  \\
\vdots & \ddots & \vdots  \\
A_{(z,1)} \, (n_{max},m_{max}) & \dots & A_{(z,N)} \, (n_{max},m_{max})
\end{pmatrix}\nonumber
\times
\\
\begin{pmatrix}
    m_1 \\
    \vdots \\
     m_N 
\end{pmatrix}
= 
\begin{pmatrix}
    b_z \, (1,0) \\
    \vdots \\
     b_z \, (n_{max},m_{max})
\end{pmatrix}.~
\end{eqnarray}
Here the matrix contains $A_{(z,i)} \, (n,m)$, the contribution from the $i$th pixel to the $n,m$ spherical harmonic mode to the magnetic field in the $z$ direction. This is then multiplied by the vector containing the magnetization of each voxel, $m_i$, to be optimized. The elements of this vector can take values of either $m_i = 1$, for the magnetization directed entirely along positive $z$, $m_i = -1$, for the magnetization directed entirely along negative $z$, or $m_i = 0$ for no magnetic material required. The result of the matrix equation in \eqref{eq:Spherical Harmonic Matrix} is the vector of total contributions from each $(n,m)$ spherical harmonic mode, $b_z (n,m)$, which are set by the user to constrain the optimization. All that remains is therefore to define the $b$ elements. The magnetic field for a MOT is well described by first order spherical harmonics, therefore these can be  targeted to produce the required linear fields. In theory this technique could be expanded to any required field, provided it can be decomposed into spherical harmonic components. This is then similar to the standard form of a linear optimization problem \cite{matousek2007understanding}, as utilized previously in \cite{schmied2009optimal}. Thus, the method employs the spherical harmonic decomposition from the process of shimming to produce a linear optimization problem to constrain the field to the desired form. This is then combined with an optimization function that seeks to maximize the contribution of the magnet strength to the field, resulting in a final placement of magnets, which produces the desired field while acting to reduce the required $M_z$.

The voxels are then defined as an assortment of readily available pre-manufactured neodymium magnets. The chosen magnets can form initial placement grids tailored to the experimental apparatus, based on their individual sizes and the required spacing between voxels (see Fig.~\ref{fig: Grid Optimisations}). Applying the optimization method then determines the required magnetization, $m_i$ of each voxel, and therefore the required positions and orientations of the magnets. For this specific setup, the initial grids were designed to allow optical access, though in theory any distribution of voxels could be defined. Examples of initial and optimized grids are shown in Fig.~\ref{fig: Grid Optimisations}. Figure~\ref{fig: Grid Optimisations}(a) shows the initial grid for cylindrical voxels of radius 6\,mm and depth 3\,mm, while Fig.~\ref{fig: Grid Optimisations}(b) illustrates the initial grid for cylindrical voxels of radius 6\,mm and depth 6\,mm. The optimized structures for each case are then shown in Figs~\ref{fig: Grid Optimisations}(c) and (d), respectively. 

The design shown in Fig.~\ref{fig: Grid Optimisations}(d) was chosen for the magnetic field generating structure, utilizing N42 strength grade neodymium magnets, and AM polymer mounts to position the magnets. The choice of final design was based on both the fidelity of the resulting field structure and the practical consideration that this design allowed the magnet rings to be centered to the CF40 viewports on the vacuum system via a simple push-fit mechanism, thus reducing the likelihood of any misalignment. The rapid prototyping provided by AM methods allows swift manufacture and application to the experiment.

The resulting fields, calculated using the derivatives of equation \eqref{eq:Greens Harmonics Full} for the optimized magnet arrangement, are compared to the target field and experimentally measured fields in Fig.~\ref{fig: Field Comparisons}. Figure \ref{fig: Field Comparisons}(a) illustrates good agreement between the target field and the numerically calculated field produced by the optimized structure of \ref{fig: Grid Optimisations}(d), for the $B_x$ and $B_z$ components along the $x$ and $z$ axes respectively. Figure \ref{fig: Field Comparisons}(b) then illustrates the agreement between the field formed by one ring (measured experimentally using a Hall probe system) and the fields produced by the same ring calculated numerically. Figure \ref{fig: Field Comparisons}(b) shows good agreement between the numerically calculated and experimentally obtained fields, with the exception of the region closest to the magnets. This small divergence most likely can be attributed to a slight misalignment of the Hall probe translation assembly relative to the magnet array. Thus, the algorithm provides a powerful method of determining a magnetic structure tailored to a given apparatus, which is capable of producing a wide variety of fields accurately.

\subsection{\label{section2.3}Vacuum system}

The central component of the vacuum system is an additively manufactured octagonal chamber carrying eight CF16 ports and two CF40 ports, as shown in Fig.~\ref{magnets}. This chamber and its production are fully described in \cite{chamberpaper}; here we give a brief summary. The mass of the chamber (excluding externally attached components) is 245\,g, considerably less than that of equivalent commercial chambers, which typically weigh $\sim$1\,kg. The chamber was additively manufactured from the aluminum alloy AlSi10Mg by a selective laser melting process \cite{lasermeltAlSi10Mg, aboulkhair2015, aboulkhair2016}. The choice of build material, and latticing of large regions of the structure to decrease the volume of solid material while preserving mechanical strength, are jointly responsible for the greatly reduced chamber mass. 

Attached to this central chamber for demonstration purposes are standard vacuum components, including a hybrid ion/NEG pump (NEXTorr D100-5), a valve for roughing and turbo-pumping of the chamber and Rb dispensers (SAES Getters). An important long-term vision is the gradual elimination of the standard vacuum components and reaching a full additive manufacture of the entire system as a single, optimized, lightweight component. Consolidating multiple modular components into one customized vacuum vessel will eliminate most of the vacuum joints present in a conventional system, improving stability while further reducing cost and SWAP parameters.  
\begin{figure}
\centering
\includegraphics[width=0.25\textwidth]{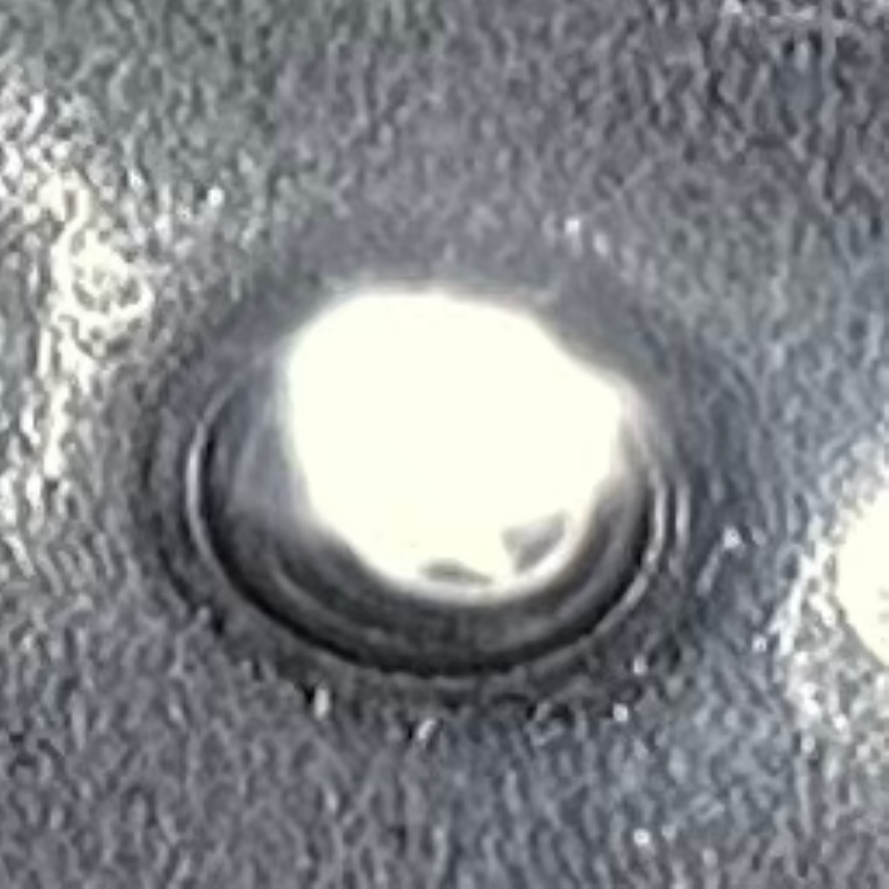} 
 \label{MOTcloud}
 \caption{Fluorescence image of the cloud of cold $^{85}$Rb atoms captured by our MOT, which was produced using the optimized components described herein.}
\end{figure}

The system was baked at 393\,K for five days, following which a pressure of $<$ 10$^{-10}$\,mbar was achieved, as measured via the ion pump current.
\section{\label{section3}Magneto-optical trapping}

The components described above were used to produce a magneto-optical trap (MOT), capturing up to 2.5$\times$10$^8$ $^{85}$Rb atoms.

The light used to form the MOT consists of three retro-reflected laser beams produced by the DFB laser systems and tapered amplifier. The cooler and the repumper beams are equally distributed via the CSPD into three optical fibers that deliver the light to the chamber. Each MOT beam contains 15\,mW of cooler light and 4.5\,mW of repumper light, with a beam a diameter of 1.2\,cm. The maximum total intensity of the six beams is $\sim$\,40\,mW/cm$^{2}$. The MOT is generated in a magnetic field gradient of 12\,G/cm provided by the ferromagnet array.

\subsection{\label{section3.1}Atom number measurement and pressure limit determination}

For an estimate of the atom number, fluorescence light from the trapped atoms was collected onto a photodiode using a plano-convex lens -- see supplementary material for details. 

Fig.~\ref{LC} shows MOT loading curves obtained for various values of Rb dispenser current from 2.20\,A to 2.75\,A. The loading curves enable a direct measurement of the pressure in the trapping region under certain conditions \cite{arpornthip, cassettari}. For each loading curve the atom number with respect to time, $N$, is fitted to the form
\begin{equation}
    N(t)= \frac{R}{L}(1-e^{-Lt})
    \label{ExpFit}
\end{equation}
where the loading rate $R$ and single-body loss rate $L$ are used as free parameters. 

This description neglects two-body and higher order loss processes, and is therefore only valid in the limit of low density of trapped atoms. In order to remain in this limit, the loading profile taken with the lowest dispenser current (2.2\,A) is used to determine an upper limit on our background pressure.

\begin{figure}
    \centering
    \includegraphics[width=0.48\textwidth]{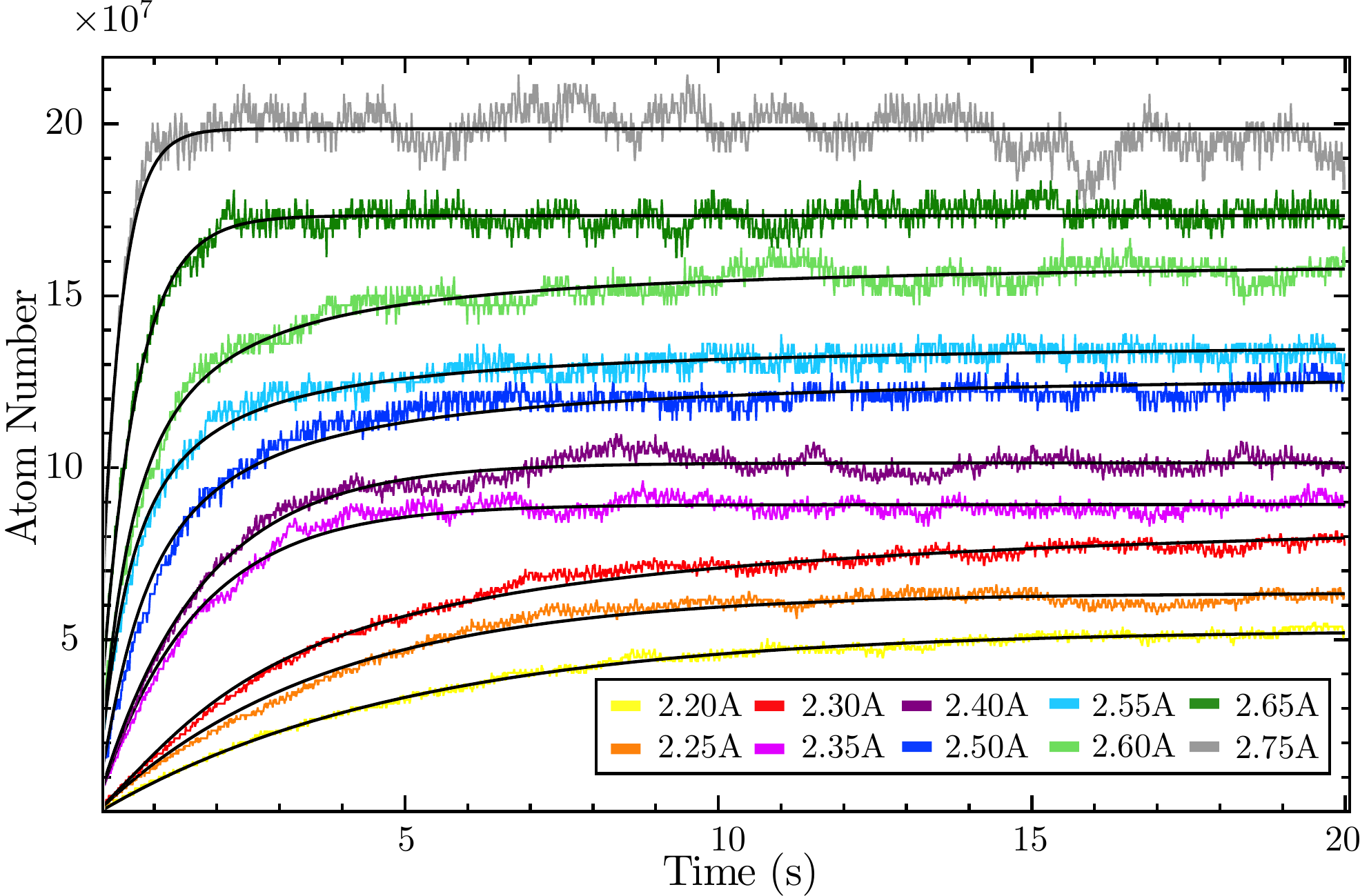} 
    \caption{MOT loading curves based on fluorescence data for various values of Rb dispenser current. Colored lines show raw data and black lines are fits to the data based on equation (\ref{ExpFit}).}
    \label{LC}
\end{figure}

A fit to this loading profile, according to equation (\ref{ExpFit}), is shown in Fig.~\ref{LC}. The extracted single-body loss rate is $L=(0.0834~\pm$\,0.0003\,s$^{-1})$. This single-body loss rate is the total loss rate resulting from collisions with all thermal background gas species present, including thermal Rb atoms. A pressure estimate based on this figure therefore represents the overall pressure in the trapping region, including the contribution from the intentionally introduced Rb atoms. Since our measurements cannot distinguish the different partial pressures of individual background species, our result represents an upper limit on the pressure of unwanted gas species, rather than a direct measurement of it.

The resulting pressure estimate depends on the assumed composition of the residual background gas. The loss coefficients per unit pressure for various common background gases have been measured in Ref.~\cite{arpornthip}. We use these coefficients to estimate the upper bound on the pressure in the trapping region under the assumption of various different dominant background gas species. The results are displayed in Table~\ref{table}. The background gas is most likely a mixture of the species listed, placing the resulting pressure limit somewhere within the range of values presented.
\begin{table}
\begin{ruledtabular}
\begin{tabular}{lc}
Species & Pressure ($\times 10^{-9}$\,mbar) \\
\hline
 H$_{2}$& 2.27\,($\pm$0.01)  \\
 He& 4.45\,($\pm$0.02)   \\
N$_{2}$ & 4.27\,($\pm$0.02) \\
\end{tabular}
\caption{\label{table}{Estimated pressures in the MOT trapping region under the assumption that the background gas consists primarily of the species indicated, corresponding to an upper limit on the non-Rb pressure.}}
\end{ruledtabular}
\end{table}
These results represent an upper limit on the pressure of non-Rb species in the chamber and are therefore consistent with the results of the ion pump current readings. While the pressure limit imposed by this measurement is much less stringent than that obtained via the ion pump current reading, it nevertheless represents an independent confirmation that the pressure is far into the high-vacuum regime. The result is particularly relevant for the future use of printed vessels in quantum technologies, opening the door for future highly complex and compact printed chambers with designs not realizable by conventional methods. 

\section{\label{section4}Conclusion}

We have demonstrated a fundamentally new approach to experimental component design that exploits the potential of AM techniques to offer greatly improved performance. AM allows direct implementation of simulation results and optimization processes.
Our results illustrate the remarkable potential of AM to facilitate experimental research in all areas currently relying on free-space optics, tailored magnetic fields, or high vacuum apparatus. The demonstrated techniques enable rapid prototyping alongside improvements in stability and substantial reductions in cost and SWAP parameters --- many component weights are reduced by 70 - 90 \% compared to standard equivalents. 
One important area of application is the field of cold atom experiments and portable quantum technologies based on magneto-optical trapping. AM components will allow widespread use of these technologies, including in field applications and space-borne experiments.

The use of AM to produce these components opens many future avenues of research. Optimum thermo-mechanical performance can be achieved via the freedom AM offers when considering material distribution, for example enabling the use of variable-density latticing \cite{becker2018space, dimopoulos2009gravitational}; optical frameworks such as the CSPD could be designed so that thermal expansion has minimal effect on the key alignment variables of the components. 
Lattice structures can also in principle be designed to isolate or damp specific frequencies of mechanical vibration \cite{khairallah2016laser}; this will be a useful feature in many experiments, as there are generally specific, narrow frequency ranges within which an experiment or device is most sensitive to environmental noise.

For AM vacuum apparatus, one promising avenue is part consolidation, in which a substantial part of a custom vacuum system could be printed as a single piece. This eliminates the overwhelming majority of the vacuum joints, further reducing SWAP parameters, increasing mechanical stability and reducing the susceptibility of the system to leaks. Another option is to exploit AM to produce high-surface area elements such as small-scale lattices or fractal surfaces. These could be coated in reactive materials to produce enhanced getter pumps for passive pumping in portable devices. 

Our demonstrated design of customized ferromagnetic arrays paves the way for progress beyond the standard magnetic field distribution used for magneto-optical trapping; systematically tailored magnetic field shapes can be produced in order to optimize selected experimental parameters, such as the total atom number or loading rate.

While AM techniques have only just started to be used in the context of quantum technologies, they hold the promise of providing a clear pathway for miniaturization and expanded functionality.


\section{Acknowledgement}
This work was supported by IUK project No.133086 and the EPSRC grants EP/R024111/1 and EP/M013294/1 and by the European Comission grant ErBeStA (no. 800942).

%

\end{document}


\title{Supplementary material for: \\ Additive manufacturing of optimised components for cold atom systems}
 \author{S H Madkhaly}
\affiliation{
School of Physics and Astronomy, University of Nottingham, University Park, Nottingham, NG7 2RD, UK}
 \affiliation{Department of Physics, Jazan University, Jazan, Kingdom of Saudi Arabia}
\author{L A Coles}
\affiliation{
Added Scientific Ltd, Unit 4, Isaac Newton Centre, Nottingham, NG7 2RH, UK
}
\author{C Morley}
\affiliation{
School of Physics and Astronomy, University of Nottingham, University Park, Nottingham, NG7 2RD, UK}
 \author{C D Colquhoun}
\affiliation{
School of Physics and Astronomy, University of Nottingham, University Park, Nottingham, NG7 2RD, UK}
 \author{T M Fromhold}
 \affiliation{
School of Physics and Astronomy, University of Nottingham, University Park, Nottingham, NG7 2RD, UK}
 \author{N Cooper}
 \email{nathan.cooper@nottingham.ac.uk}
\affiliation{
School of Physics and Astronomy, University of Nottingham, University Park, Nottingham, NG7 2RD, UK}
 \author{L Hackerm\"{u}ller}
 \email{lucia.hackermuller@nottingham.ac.uk}
\affiliation{
School of Physics and Astronomy, University of Nottingham, University Park, Nottingham, NG7 2RD, UK}

\maketitle

\newpage
\section*{Material selection}

\noindent Different 3D printing materials were tested to ensure their relevance and compatibility with optical systems. Formlabs `Rigid Resin' was chosen as the build material for the DFB housing, lens tubes mounts, and CSPD on the basis of its low coefficient of thermal expansion and high elastic modulus, which offer improved alignment stability when compared to alternative build materials. Table \ref{table} details physical and thermal properties of the selection of build materials considered. The material from which our system parts are 3D printed `rigid resin' is shaded in grey.

\begin{table}
\begin{ruledtabular}
\centering
\begin{tabular}{P{2.75cm}P{2cm}P{2.75cm}P{2.75cm}P{3cm}P{1.75cm}}
Material&
Supplier&
Elastic Modulus (GPa)&
Glass transition Temp. ($^{\circ}$C)&
Thermal expansion coefficient ($\alpha$)($\mu$m/m)/\, $^{\circ}$C&
Reference \\ 
\colrule
Polycarbonate (PC) & Ultimaker & 2.13 & 147 & 69 & \cite{ultimaker, simplify3d, CantrellJasonT2017Ecot, kim2017experimental} \\ \hline
Polylactic\,Acid (PLA) & Ultimaker & 2.35  & 60 & 68 & \cite{ultimaker,simplify3d} \\\hline
Acrylonitrile Butadiene Styrene (ABS) & Ultimaker & 1.62  & 105 & 90 & \cite{ultimaker, simplify3d, CantrellJasonT2017Ecot, kim2017experimental} \\\hline
Co-polyester (CPE) &  Ultimaker & 1.54 & 82 & 70 & \cite{ultimaker,simplify3d}\\\hline
\rowcolor[gray]{0.96}
Photopolymer (rigid)\,resin (White) & Formlabs & 4.1 & 88

(heat deflection) & 53 & \cite{rigidresin}\\\hline 
Photopolymer (tough)\,resin (Green) & Formlabs & 2.7 & 45

(heat deflection) & 119.4 & \cite{toughresin}\\
\end{tabular}
\end{ruledtabular}
\caption{Physical and thermal properties of candidate AM build materials for the CSPD\cite{ABSSS,PLAAA,CPEEEE}.}
\label{table}
\end{table}

\section*{Optical pumping effects}

\begin{figure}[h!]
    \centering
    \includegraphics[width=0.9\textwidth]{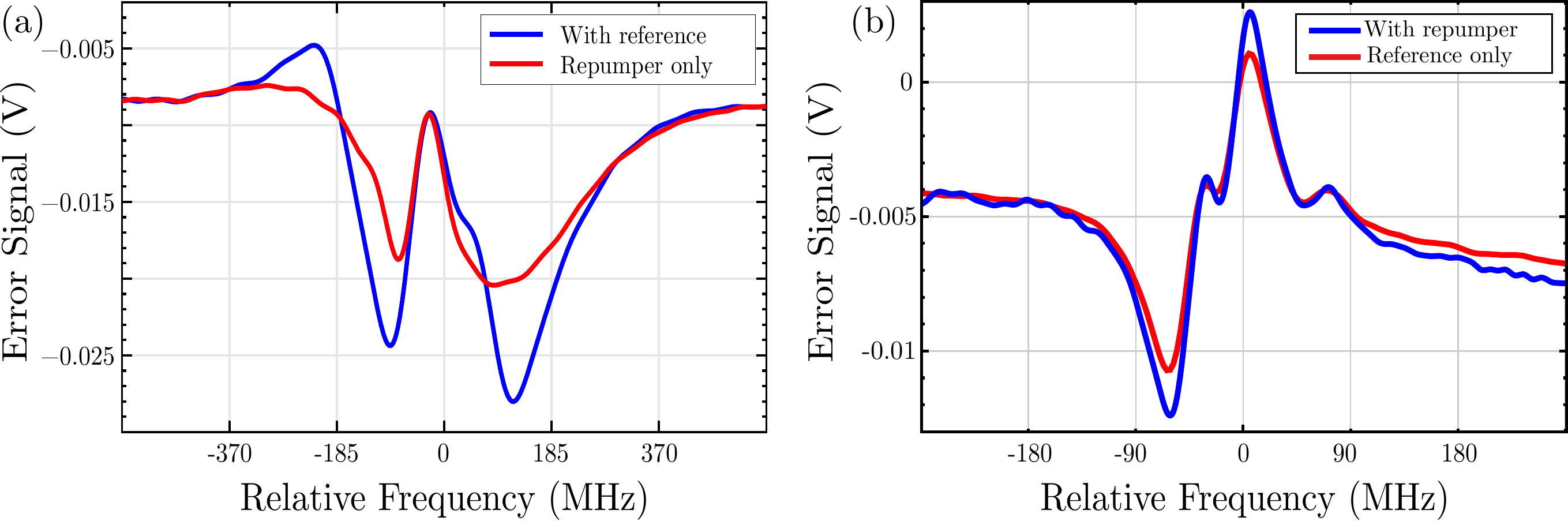}
    \caption{Error signals resulting from saturated absorption spectroscopy (with laser current modulation and phase-sensitive detection) of the $^{85}$Rb D2 line $|F=2\rangle \rightarrow |F=1,2,3\rangle$ `repumper' transition (a) and $^{85}$Rb $|F=3\rangle \rightarrow |F=2,3,4\rangle$ `reference' transition (b). The red curves represent the signals generated when only the laser performing the spectroscopy was present in the vapour cell, while the blue lines represent the signals obtained when a beam resonant with the other transition spatially overlapped with this beam in the same vapour cell. As can be seen from the figures, optical pumping effects result in an enhancement of the error signal amplitude --- particularly for the laser addressing the repumper transition.}
    \label{fig3}
\end{figure}

\noindent As described in Section 2.2 of the main article, the use of overlapping repumper and reference beams in the spectroscopy cell of the CSPD results in a modification of the spectroscopy signals. This means that accurate frequency stabilization for either laser requires both lasers to be frequency stabilized. However, it also enhances the amplitude of the spectroscopic and stabilization signals generated by the CSPD, improving signal to noise and consequently enabling more stable locking --- particularly for the repumper laser. Figure \ref{fig3} shows the `error signals' used for feedback stabilization of the laser frequencies, generated by modulating the laser current and then combining the spectroscopic and current modulation signals via an analog multiplier. The amplitude enhancement resulting from the presence of the second laser on the same spatial path can clearly be seen in each case, and results in more pronounced locking signals and steeper lock-points. 

\section*{Hybrid magnetic field generation --- detailed derivation}

\noindent In section II C of the main article, we claim that augmentation of coils with permanent magnets can reduce the time-averaged power requirements for magnetic field generation by a factor equal to $1/(1-T_M)$ compared to using coils alone, where $T_M$ is the fraction of the experimental cycle time for which the magnetic field must be active. Here, we provide a full derivation of this result.

\noindent The strength of the magnetic field generated by a coil is proportional to the current through it, $I$, while the power dissipated in that coil is equal to $I^2 R$, with $R$ the coil's resistance. We neglect inductive effects because in practice experimental cycle times are usually sufficiently long to make them negligible as a source of power consumption, while from a purely theoretical perspective they do not necessarily have to result in a net energy loss from an appropriately designed system. It follows that the power dissipation in a set of coils is equal to $R\overline{I^2}$, and hence also to $C\overline{B^2}$, where $B$ is a scalar proportional to the magnetic field strength produced by the coils and $C$ is a constant coefficient dependent on the exact system parameters. 

\noindent Let us now define a fraction, $F$, of the required magnetic field that is to be produced by a ferromagnetic array, with the remaining fraction, $1-F$, being produced by coils. We assume that the fields are of the same form and consider only their relative magnitudes. The power consumption while the field is to be on is now equal to $C(1-F)^2B^2$. However, while the field is required to be off the coils must be used to cancel the field component of the permanent magnets by running current in the opposite direction. The power consumption for this process is clearly now equal to $CF^2B^2$. 

\noindent We now define the fraction of the experimental cycle for which the MOT fields are to be active as $T_M$. From this we see that the time-averaged power consumption $\overline{P}$ is given by
\begin{equation}
    \overline{P} = T_M C (1-F)^2 B^2 + C (1-T_M) F^2 B^2.
    \label{power}
\end{equation}

\noindent In order to find the value of $F$ that mimimises time-averaged power consumption, we set the first derivative of $\overline{P}$ with respect to $F$ equal to zero. This gives
\begin{equation}
    C B^2 \left[ -2 T_M (1-F) + 2F (1-T_M) \right] = 0,
\end{equation}
the solution to which is $F=T_M$. 

\noindent Using a standard approach where the entire required field is generated by the coils, the time-averaged power consumption would simply be equal to $T_M C B^2$. Setting $F=T_M$ in equation (\ref{power}) and dividing through by this value, we find that the use of permanent magnets to create part of the MOT field reduces time-averaged power consumption by a factor of $1/(1-T_M)$.

\section*{Atom number estimation}
\noindent The MOT atom number $N$ is calculated from the photodiode voltage $V_\mathrm{PD}$ using:
\begin{equation}
    N = \frac{4\pi\lambda\alpha V_\mathrm{PD}}{hc\Omega\gamma_\mathrm{sc}s},
\end{equation}
\noindent where $\gamma_\mathrm{sc}$ denotes the photon scattering rate of the MOT atoms, $\alpha$ is the magnification factor of the lenses used to capture the MOT fluorescence, and $s$ is the sensitivity of the photodiode to 780\,nm light. Loading curves (see FIG.9 in the main paper) were obtained for various values of Rb dispenser current from 2.20\,A (minimum current value at which an atomic cloud can be formed) to 2.75\,A.